# Seismogenic Mobile Network

Manana Kachakhidze[1,2], Nino Kachakhidze-Murphy[3], Badri Kvitia[3], Giorgi Khazaradze[4]

[1]LEPL Institute "OPTIKA", Tbilisi, Georgia
[2]Georgian Technical university, Tbilisi, Georgia
[3]Ivane Javakhishvili Tbilisi State University, Tbilisi, Georgia
[4]Group RISKNAT, Institute Geomodels, University of Barcelona, Barcelona, Spain

**Abastract**

It is known that VLF / LF radiation existing prior to large earthquakes is recorded in several seismically active countries of the world. The networks of this radiation consist of stationary transmitters and receivers. However, there are cases of large earthquakes, when existing networks can not fix relevant EM radiation.

In the presented paper we will focus on the optimal option of network arrangemnet of the VLF / LF electromagnetic radiation before the earthquakes as a earthquakes precursor.

The paper discusses the new possibility of arrangement of VLF / LF network based on certain physical considerations, which is relatively simplified and completely different from the existing networks. It will increase to the limit the number of earthquakes fixed with the relevant EM emissions which is the aim of the new network arrangement.

**Introduction**

Studies of earthquake problems in the world were especially intensified from the second half of the last century, since alongside with theoretical studies it became possible to carry out high level laboratory and satellite experiments. Thanks to them during earthquake preparation process various anomalous changes of geophysical fields have been revealed in lithosphere as well as in atmosphere and ionosphere. These different geophysical phenomena may accompany earthquake preparation process and expose themselves several months, weeks or days prior to earthquakes.

Among the anomalous geophysical phenomena preceeding earthquake the specific attention is attributed to earth electromagnetic emissions before earthquakes, because correlation between seismic activity and disturbances in radiobroadcasts is revealed by scientists.

We meet rather diverse and interesting scientific papers, on the basis of groundbased and satellite data of earth VLF/LF and ULF electromagnetic (EM) emissions observed during earthquake preparation period (Bahat, et al., 2005; Biagi, et al.,1999, 2007, 2009, 2011, 2012, 2013, 2014, 2019; Bleier, et al., 2010 ; Boudjada et al., 2010, 2017 ; Contoyiannis et al., 2014; Denisenko et al., 2008; Dolea et al., 2015; Eftaxias, et al., 2003, 2009; Hayakawa and Fujinawa, 1994; Hayakawa, 1999; Hayakawa and Molchanov, 2002; Hayakawa, et all, 2006; Hayakawa and Molchanov, 2007; Maggipinto et al, 2015; Mognaschi,et al., 1993; Moldovan, et al., 2015; Molchanov et al., 1998; Onishi, et al., 2011; Politis,et al.,2020 ; Potirakis, et al., 2015; Rozhnoi, et al., 2009; Uyeda, et al., 2009; Yoshino,1991; Zhang et al., 2012).

The various networks were arranged to collect VLF/LF radio signals in order to study of electromagnetic field variations associated with seismogenic processes.

Since 2000 Japanese researchers have been able to rely on VLF radio network (the Pacific network) which has seven receivers to measure the intensity and phase of VLF radio signals from two different transmitters. On February 2002, in the framework of a scientific cooperation among Japanese, Russian and Italian teams, a receiver was placed into operation in the Department of Physics at the University of Bari, which marked the beginning of the development of a Eroapean network (Maggipinto et al., 2013).

Since 2009 a network of VLF (20 - 60 kHz) and LF (150 - 300 kHz) radio receivers is operating in Europe in order to study the disturbances produced by the earthquakes on the propagation of these signals.

VLF radio signals lie in the 10 - 60 kHz frequency band. These radio signals are used for worldwide navigation support, time signals and for military purposes. They are propagated in the earth-ionosphere wave-guide mode along great circle propagation paths. So, their propagation is strongly affected by the ionosphere conditions. LF signals lie in 150 - 300 kHz frequency band. They are used for long way broadcasting by the few (this type of broadcasting is going into disuse) transmitters located in the world. These radio signals are characterized by the ground wave and the sky wave propagation modes. The first generates a stable signal that propagates in the channel earth-troposphere and is affected by the surface ground and troposphere condition. The second instead gives rise to a signal which varies greatly between day and night, and between summer and winter, and which propagates using the lower ionosphere as a reflector; its propagation is mainly affected by the ionosphere condition, particularly in the zone located in the middle of the transmitter-receiver path. The propagation of the VLF/LF radio signals is affected by different factors such as the meteorological condition, the solar bursts and the geo-magnetic activity. At the same time, variations of some parameters in the ground, in the atmosphere and in the ionosphere occurring during the preparatory phase of earthquakes can produce disturbances in the above mentioned signals. As already reported by many previous studies the disturbances are classified as anomalies and different methods of analysis as the residual dA/ dP, the terminator time TT, the Wavelet spectra and the Principal Component Analysis have been used (Biagi et al., 2012).

The research of seismic effects on the VLF/LF radio signals is based on the spotting of disturbances on the data. In this framework, a first and fundamental step is the identification of possible disturbances related to causes different from the seismicity (Biagi et al., 2013). If after analysing the radio data of each day in order to reveal anomalies and the different causes of the revealed anomalies will be excluded by the researchers, a seismic effect should be realistic (Biagi et al., 2011, Biagi, 2019).

It is true that satellite and terrestrial networks record electromagnetic radiation before an earthquake but there are instances of strong earthquakes when radiation can not be detected (Biagi et al., 2011; Biagi, 2019).

This fact raises doubts among the scientists that EM emissions may not always exist during the earthquake preparation period, which of course, contradicts the fact of the EM origination during the cracks forming process. Although, this fact is proven for a long time not only theoretically but experimentally too (Bleier et al., 2010; Contoyiannis et al., 2015; Eftaxias, et al., 2002, 2007a, 2007b, 2008, 2009, 2010. Freund, et al., 2006; Gershenzon, N., Bambakidis, G., 2001 ; Hadjicontis, et al, 2007; Ikeya and Takaki, 1996; Papadopoulos, et al., 2010; Politis, et al., 2020; Potirakis, et al., 2013, 2015; Rabinovitch,et al., 2007; Schwingenschuh, et al., 2011; Yoshida, et al., 1997).

Nevertheless, the role of INFREP (as well as Pacific network and satellite data) is immense in the study of EM emissions existing before earthquakes. They were the first and necessary step. Without them the richest scientific researches would not be created which have expanded the scope of earthquake problem studies and the possibilities for predicting an earthquake have become real.

Of course, we agree with the position of scientists who have been investigating electromagnetic emissions in the nature and laboratories for years and based on research conclude that the VLF/VHF electromagnetic precursors do exist and that the development of suitable observational techniques and analysis methods is a promising research direction for EQ precursors study (Eftaxias et al., 2003).

**Discussion**

It is obvious, that the geophysical field, which can be considered as earthquake precursor, has to precisely express the geological model of fault origination in the focus (Mjiachkin,1978) and in connect with it, should analytically describe the most difficult process from the beginning of micro cracks appearing up to main fault formation and restore the equilibrium in the focus.

Discussing the problem of earthquakes, the scientist has no right to solve the task by allowing certain boundary conditions, because it will not give us a real result, especially the possibility of predicting an earthquake.

Experimental studies in the direction of searching of VLF/LF EM radiation existent before earthquake have shown that: 1) in the period of large earthquake preparation, noted radiation begins a few weeks before the earthquake; 2) for the spectrum of existing electromagnetic emissions the following sequence of frequencies is characteristic: MHz, kHz. 3) the both these emissions from the beginng up to end are accompanied by ULF radiation; 4) in most cases, a few days before the earthquake, so-called "silence" of emissions takes place. During "silence" electromagnetic field radiation almost does not exist or it is reduced; 5) "silence" of EM emissions is followed by earthquake (Eftaxias et al., 2009; Eftaxias et al., 2013; Papadopoulos, et al., 2010).

The existence of this type of VLF / LF field in the epicenter area during the earthquake preparation period and its tendency to change, indicates that: i) The emitted body of VLF / LF EM emissions should be in the incoing earthquake focus, ii) The character of changes of the VLF / LF EM emissions existing in the epicentral area should be caused by changes of length of this body;

Based on these considerations, the model of generation of EM emissions fixed before earthquake (Kachakhidze et al., 2015) and later, based on the INFREP retrospective data, methods of large earthquake prediction have been developed (Kachakhidze et al., 2019).
It turned out that:

1) About few dozen days prior to the earthquake, it is possible to separate a continuous active frequency channel;
2) By the active channel frequency, about few dozen days before the earthquake, it is possible to determine the length of "cracked strip" on which the process of cracks origination is going on actively and ultimately the main fault is formed;
3) By the length of the "cracked strip", it is possible to determine magnitude of incoming earthquake with certain accuracy about few dozen days prior to the earthquake;
4) After the active frequency channel detection, it already is possible to determine the future earthquake epicenter with certain accuracy;
5) In order to short-term prediction of a large earthquake, it is recommended to begin the frequency data careful monitoring from the starting moment of the avalanche-unstable process of fault formation and keep an eye on the process dynamics;
6) In case of monitoring of electromagnetic emissions existent before earthquake, it is possible, step-by-step, to make about few dozen as well as 2-days short-term prediction of incoming earthquake;
7) Based on the proposed method, it is easy to separate the foreshock and aftershock series from the main shock;

Finally, we may conclude that EM emissions turned really out to be the unique precursor, which is capable of large earthquake short-term prediction.

The results of this work confirm above mentioned theoretical model that during the earthquake preparation period, the direct cause of generation of the VLF /LF frequencies electromagnetic field is origination and formation of the fault in the earthquake focus (Kachakhidze et al., 2019).

Because the fault, existing in focus, is the radiating body of the VLF / LF electromagnetic field (Kachakhidze et al., 2015), it is obvious, that in order to record the radiated field data, the receiver device of this field must be located at place so, that it can directly see the field radiated from the fault.

In order to strengthen the model (Kachakhidze et al., 2015) of generation of EM emissions detected prior to earthquakes with experimental data and to make prognostic conclusions, in the above study we used the exactly that retrospective data of the INFREP network which most likely had to satisfy the obligatory condition, necessary for fixing of reliable EM emissions data, relevant to earthquakes. Namely, the appropriate receiver must directly "see" and record the electromagnetic emissions.

That is why the discussed Crete earthquake was purposely selected (Kachakhidze et al., 2019). Research on these earthquake data was conditioned by the fact that VLF/LF EM emissions receiver is installed in the Crete territory where exactly earthquake occurred.

Although the INFREP network receivers get the data from the transmitter, it is possible to assume that the receiver in Crete, besides the data getting from transmitter, recoreded the information directly from the radiating body as it "saw" this field.

The study confirmed that, the receiver really recorded such data and based on them it was possible to describe the Crete large earthquake preparation process with high precision, or we had a possibility of description of fault formation process (with quite high accuracy) appeared in the focus and subsequently, it afforded the opportunity to create the large earthquake short – term prediction methods (Kachakhidze et al., 2019).

Of course, the most important for earthquake forecasting is to record the parameters of that relevat geophysical field with high accuracy, which is responsible for forecasting.

Because of certain works (Eftaxias et al., 2002, 2009; 2010; Biagi et al., 2009, 2013; Kachakhidze et al., 2015), EM emissions prior to earthquake, turned out like such field, there is no doubt that the receiver of EM field really must "see" and fix the characteristic parameters of the electromagnetic field originated in the focus of the earthquake during earthquake preparation process.

In addition, for recording of M≥5.0 magnitude earthquakes, the receiver must record VLF /LF radiation in (102 kHz-0.377 kHz) frequency diapason (Kachakhidze et ail., 2015). This is one of the essential and necessary requirements, that a modern VLF/LF EM emissions recorder network must satisfy.

Today existing VLF/LF networks, which consist mainly of stationary transmitters and receivers, record EM emissions spectrum in only 20-60 kHz (VLF) and 150-300 kHz (LF) frequency diapason through 10 channels (for example, INFREP). This means that there exists an opportunity to fix the earthquakes only with certain magnitudes (Biagi et al., 2014; Biagi et al., 2019).

It is known that large earthquake is preparing for quite a long time, which means that the process of tectonic stress accumulation takes long time in any seismic active region (or country).

Due to it, anomalous sites of tectonic stress of this territory, must be under geodetic observations for a long time (Banks et al.,1997; Reilinger, et al., 2006; Khamidov, 2017; Sokhadze et al., 2018;; Elshin and Tronin, 2020).

In collaboration with USA scientists, these observations are carried out since the end of the last century in Transcaucasian countries and in Georgia among them too (Reilinger et al., 2006; Sokhadze et al., 2018).

Based on these observations, the region (or country) is studied by geodetic, geologic and seismic views.

Let us consider the research conducted in Georgia for example. In this study authors present and interpret new GPS observations made during the period 2008 through 2016 for 21 survey-mode sites and 9 continuous stations. The GPS observations are primarily aligned along two roughly range-perpendicular profiles that cross the Lesser–Greater Caucasus boundary zone. Objectives of searching are to determine the rate of active deformation across these two segments of the boundary and use the observed deformation and elastic fault models to constrain the locations and character of active structures in this portion of the Arabia–Eurasia collision zone (Fig.1) (Sokhadze et al., 2018).

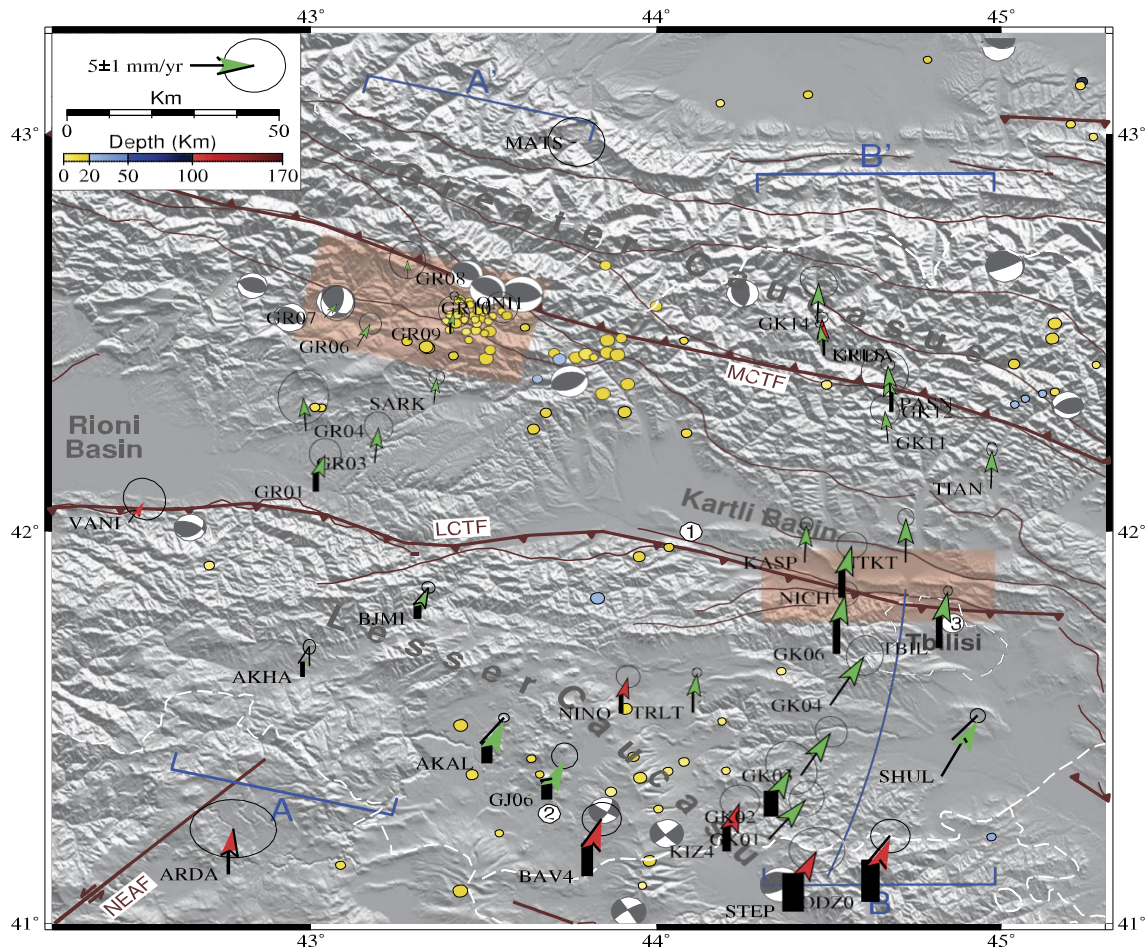

Fig. 1

It turned out that convergence between the Lesser and Greater Caucasus along the eastern Rioni Basin is primarily accommodated on a north-dipping fault system along the southern margin of the Greater Caucasus. The geodetically estimated location and dip of the model fault is consistent with the location and fault parameters reported for the 1991 Mw=6.9 Racha earthquake, which occurred on the MCTF at depth (Sokhadze et al., 2018).

In contrast, principal convergence between the Lesser and Greater Caucasus across the Tbilisi segment, immediately east of the Rioni segment, occurs along the northern boundary of the Lesser Caucasus. Best-fit fault models involve convergence on a north-or south-dipping thrust fault located near the northern edge of the Lesser Caucasus, approximately 50–70 km south of the MCTF, and near Tbilisi. Authors suggest that the southward offset of the convergence zone is related to the incipient collision between the Lesser and Greater Caucasus. Ongoing seismic, geodetic and tectonic investigations promise to better characterize the source of observed strain and, accordingly, implications for seismic hazards and the tectonic evolution of the LC-GC collision zone (Sokhadze et al., 2018).

Such searching let us separate the site (or sites) in a given region where tectonic stress is gathered and relevant anomalies are revealed several month (years) before of earthquake. In the case of Georgia, according to the geological interpretation, because of arising of tectonic anomaly near Tbilisi, it is necessary to reevaluate the seismic hazards in this area.

The facts that earthquakes occur in areas of accumulation of tectonic stress and mainly these areas are created in any region (country) a few months or years before the earthquake (Sokhadze et al., 2018; Elshin and Tronin, 2020), gives us the opportunity to think about modernization of the today being networks of electromagnetic radiation existent before the earthquake. This means that, in order to predict an earthquake, the network of mobile receivers of VLF / LF electromagnetic emissions, without transmitters, must be purposefully installed only for specific tectonic anomaly of the region (or country). Such a network may be called a "seismogenic mobile network".

It should be noted that, before arrangement of seismogenic mobile network, after numeric estimation of tectonic stress area (even in the case of its constantly changing) it is possible in advance, step by step to determine, approximate value of the incoming earthquake magnitude by the Dobrovolsky formula (Dobrovolsky, et al., 1979).

$$R = 10^{0.43\,M} \quad km$$

where $R$ is the „strain radius".

In such a case, by monitoring of incoming earthquake magnitude, it is also possible step by step to estimate approximate changes of the fault lengths in the focus of incoming earthquake (Kachakhidze et al., 2015). Therefore the "area of vision" of the seismogenic mobile network of VLF/LF EM emissions existent prior to the earthquake, in the case of different magnitudes, obviously, will be different.

These works can be carried out at any early stage of tectonic stress accumulation and in order to control the process of gathering of tectonic tension, be continued for quite a long time.

In case of reaching a certain limit of tectonic tension, when it becomes necessary to involve earthquake prediction works and accordingly, arranging the mobile seismogenic network on a specific seismogenic area, above mentioned works allow us to determine the required number of EM emissions receivers and precisely select locality where the best reception of VLF / LF EM radiation from the earthquake focus will be guaranteed by the mobile receivers installed in these places.

The arragement of stationary networks throughout the whole territories of the region (or country) is justified for areas where geodetic surveys are not conducted in order to detect geotectonic anomalies.In addition, it should also be noted that stationary EM emissions networks, which except for receivers consists of transmitters too, are quite expensive.

Thus, in order to completely recording of electromagnetic emissions existent before earthquake, it is necessary first of all to conduct geodetic surveys to detect tectonic anomalies in any region (or country), and then to arrange a seismogenic mobile network at the specific sites of tectonic anomalies.

„Seismogenic mobile network" has some advantages compared to the today existing stationary networks:
1) "Seismogenic mobile network" does not need having of transmitters and consists of only EM radiation mobile receivers;
2) It is easy to move the mobile seismogenic network to any new survey site due to tectonic anomalies locations;
3) Mobile seismogenic network is convenient because in case of each concrete incoming earthquake, it is possible to change the number of receivers because of the specific fault length;
4) It must be keep in mind that the perfect recording of radiated electromagnetic emissions depends on the direction of the main fault, generated in the earthquake focus, towards the receiver, which may be determined with certain accuracy during long time observation;
5) According to the searching, the active fault, where the earthquake preparing process is going on, may be fixed about some dozen days before earthquake and determined fault length in the focus with certain accuracy (Kachakhidze et al., 2015, 2019). It means that in order to get complete data of electromagnetic emissions, the interconfiguration and distances from each other of electromagnetic emissions receivers of mobile seismogenic network, may be choosen optimal. It should be noted that it is desirable to determine approximate location of the earthquake focus in advance based on the data of tectonic anomalies (Elshin and Tronin, 2020).
6) Receivers, since they should be located mainly in the probable epicenter of the upcoming earthquake, will fix the full spectrum of electromagnetic emissions (with some accuracy, of course) which will be radiated by fault; To control the full spectrum of radiation means to measure EM emisiions in (102 kHz-0.377 kHz) frequancy diapason necessary to predict M≥5.0 magnitude earthquakes.
7) Using of mobile network, including initial geodetic searching, will become earthquake

prediction work much simpler and cheaper.

**Conclusion:**

In the presented work, the considerations about expediency of arrangement of mobile seismogenic network of VLF/LF electromagnetic emissions, existent before earthquake, are offered.

In any seismoactive region (country), where the general tectonic stress picture is studied in advance by geodesic searching, the following possibilities arise:

1) To arrange purposefully the mobile seismogenic network of EM emissions in areas of pre-identified specific geotectonic anomalies according to a preliminary rough calculation of the expected magnitude of the incoming earthquake.
2) To carry out the high precise measurements of VLF/LF EM emissions for earthquakes with M≥5.0 magnitudes in (102 kHz-0.377 kHz) frequency diapazon .
3) Mobile seismogenic networks will be organized with much smaller financial cost, which is due to the fact that the seismogenic mobile network consists only of EM emissions receivers, the location of which can be easily changed according to changing of location of geotectonic anomalies; This will allow us to optimally use human resources as well as the necessary material and technical base.

As for already exsting networks, in order to improve them, according to above described principle of arranging of the seismogenic networks, by our view, it will be enough to add several mobile receivers to them.

REFERENCES:

1. Banks, C.J., Robinson, A.G., Williams, M.P. Structure and regional tectonics of the Achara–Trialet fold belt and the adjacent Rioni and Karli foreland basins, Republic of Georgia. In: Robinson, A.G. (Ed.), Regional and Petroleum Geology of the Black Sea and Surrounding Region, pp. 331–346, 1997.

2. Bahat, D., Rabinovitch, A., and Frid, V.: Tensile fracturing in rocks: Tectonofractographic and Electromagnetic Radiation Methods, Springer Verlag, Berlin, 570 pp., 2005.

3. Biagi, P.F. Seismic effects on LF radiowaves. In: Hayakawa, M. (Ed.), Atmospheric and Ionospheric Electromagnetic Phenomena Associated with Earthquakes. TERRAPUB, Tokyo, pp. 535–542, 1999.

4. Biagi, P. F., Castellana, L., Maggipinto, T., Maggipinto, G., Minafra, A., Ermini, A., Capozzi, V., Perna, G., Solovieva, M., Rozhnoi, A., Molchanov, O. A., and Hayakawa, M. Decrease in the electric intensity of VLF/LF radio signals and possible connections. Nat. Hazards Earth Syst. Sci., 7, 423–430, 2007. www.nat-hazards-earth-syst-sci.net/7/423/2007/ .

5. Biagi, P.F., Castellana, L., Maggipinto, T., Loiacono, D., Schiavulli, L., Ligonzo, T., Fiore,M., Suciu, E., Ermini, A. A pre seismic radio anomaly revealed in the area where the Abruzzo earthquake (M = 6.3) occurred on 6 April 2009. Nat. Hazards Earth Syst. Sci. 9, 1551–1556. 2009. http://dx.doi.org/10.5194/nhess-9-1551-2009 .

6. Biagi, P.F., Maggipinto,T., Righetti,F., Loiacono,D., Schiavulli,L., Ligonzo,T., Ermini,A., Moldovan,I.A., Moldovan, A.S., Buyuksarac, A, Silva,H.G., Bezzeghoud,M. and Contadakis, M.E. The European VLF/LF radio network to search for earthquake precursors: setting up and natural/man-made disturbances. Nat. Hazards Earth Syst. Sci., 11, 333–341, 2011. www.nat-hazards-earth-syst-sci.net/11/333/2011/ doi:10.5194/nhess-11-333-2011 .

7. Biagi, Pier Francesco., Righetti, Flavia., Maggipinto, Tommaso., Schiavulli, Luigi., Ligonzo,Teresa., Ermini, Anita., Moldovan, Iren Adelina., Moldovan, Adrian Septimiu., Silva, Hugo Gonçalves., Bezzeghoud, Mourad., Contadakis, Michael E., Arabelos, Dimitrios N., Xenos, Thomas D., Buyuksarac, Aydin. Anomalies Observed in VLF and LF Radio Signals on the Occasion of the Western Turkey Earthquake (Mw = 5.7) on May 19, 2011. International Journal of Geosciences,3, 856-865, 2012. http://dx.doi.org/10.4236/ijg.2012.324086. Published Online. September 2012 http://www.SciRP.org/journal/ijg .

8.  Biagi, P.F., Magippinto, T., Schiavulli, L., Ligonzo, T., Ermini, A. European Network for collecting VLF/LF radio signals (D5.1a). DPC- INGV - S3 Project. "Short Term Earthquake prediction and preparation", 2013.

9.  Biagi, P.F., Maggipinto, T., Ermini, A. The European VLF/LF radio network: current status. Springer. Acta Geod Geophys. 2014. DOI 10.1007/s40328-014-0089-x .

10. Biagi, P. F., Colella, R., Schiavulli, L., Ermini, A., Boudjada, M., Eichelberger, H., Schwingenschuh, K., Katzis, K., Contadakis, M. E., Skeberis, C., Moldovan, I. A., Bezzeghoud, M. The INFREP Network: Present Situation and Recent Results. Scientific Research Publishing. Open Journal of Earthquake Research, 2019, 8, 101-115. 2019. http://www.scirp.org/journal/ojer

11. Bleier, T., Dunson, C., Maniscalco, M., Bryant, N., Bambery, R., Freund, F., 2009.Investigation of ULF magnetic pulsations, air conductivity changes, and infra red signatures associated with the 30 October Alum Rock M5.4 earthquake. Nat. Hazards Earth Syst. Sci. 9, 585–603. http://dx.doi.org/10.5194/nhess-9-585-2009 .

12. Bleier, T., Dunson, C., Alvarez, C., Freund, F., and Dahlgren, R. Correlation of pre-earthquake electromagnetic signals with laboratory and field rock experiments. Nat. Hazards Earth Syst. Sci., 10, 1965–1975, 2010. www.nat-hazards-earth-syst-sci.net/10/1965/2010/ doi:10.5194/nhess-10-1965-2010

13. Boudjada, M. Y., Schwingenschuh, K., D¨oller, R., Rohznoi, A., Parrot, M., Biagi, P. F., Galopeau, P. H. M., Solovieva, M., Molchanov, O., Biernat, H. K., Stangl, G., Lammer, H., Moldovan, I., Voller, W., and Ampferer, M. Decrease of VLF transmitter signal and Chorus-whistler waves before l'Aquila earthquake occurrence. Nat. Hazards Earth Syst. Sci., 10, 1487–1494, 2010. www.nat-hazards-earth-syst-sci.net/10/1487/2010/ . doi:10.5194/nhess-10-1487-2010 .

14. Boudjada, M.Y., Biagi, P.F., Al-Haddad, E., Galopeau, P.H.M., Besser, B., Wolbang, D., Prattes, G., Eichelberger, H., Stangl, G., Parrot, M., Schwingenschuh, K. Reception conditions of low frequency (LF) transmitter signals onboard DEMETER micro-satellite. Physics and Chemistry of the Earth,Volume 102, pages 70-79. December 2017. https://doi.org/10.1016/j.pce.2016.07.006

15. Contoyiannis, Y., Potirakis, S.M., Kopanas, J., Antonopoulos, G., Koulouras, G., Eftaxias, K., Nomicos, C. On the recent seismic activity at Kefalonia island (Greece): manifestations of an earth system in critical state. Geophysics, 1–12, 2014.arXiv: 1401.7458.

16. Denisenko,V. V., Boudjada, M. Y., Horn, M., Pomozov, E. V., Biernat, H. K., Schwingenschuh, K., Lammer, H., Prattes, G., and Cristea, E. Ionospheric conductivity effects on electrostatic field penetration into the ionosphere. Nat. Hazards Earth Syst. Sci., 8, 1009–1017, 2008. www.nat-hazards-earth-syst-sci.net/8/1009/2008 /

17. Dobrovolsky, I.P., S.I. Zubkov and V.I. Miachkin. Estimation of the Size of Earthquake Preparation Zones, Pageoph, Vol. 117, 1025-1026, 1979. Birkhäuser Verlag, Basel.

18. Dolea, Paul., Cristea, Octavian., Vladut Dascal, Paul., Moldovan, Iren-Adelina., Biagi, Pier Francesco. Aspects regarding the use of the INFREP network for identifying possible seismic precursors. Physics and Chemistry of the Earth 85–86, 34–43, 2015. http://dx.doi.org/10.1016/j.pce.2015.05.010

19. Eftaxias, K., Kapiris,P., Dologlou, E., Kopanas, J., Bogris, N., Antonopoulos, G., Peratzakis, A. and Hadjicontis. V. EM anomalies before the Kozani earthquake: A study of their behavior through laboratory experiments. Geophysical Research Letters, VOL. 29, NO. 8, 1228, 10.1029/2001GL013786, 2002.

20. Eftaxias, K., P. Kapiris, J. Polygiannakis, A. Peratzakis, J. Kopanas, G. Antonopoulos, and D. Rigas. Experience of short term earthquake precursors with VLF–VHF electromagnetic emissions. Natural Hazards and Earth System Sciences, 3, 217–228, 2003.

21. Eftaxias, K., Sgrigna, V., Chelidze, T. Mechanical and electromagnetic phenomena accompanying preseismic deformation: from laboratory to geophysical scale. Tectonophysics 431, 1–301, 2007a

22. Eftaxias, K., Panin, V., Deryugin, Y. Evolution EM-signals before earthquakemand during laboratory test of rocks. Tectonophysics 431, 273–300, 2007b.

23. Eftaxias, K., Contoyiannis, Y., Balasis, G., Karamanos, K., Kopanas, J., Antonopoulos, G., Koulouras, G. and Nomicos, C. Evidence of fractional-Brownian-motion-type asperity model for earthquake generation in candidate pre-seismic electromagnetic emissions. Nat. Hazards Earth Syst. Sci., 8, 657–669, 2008. www.nat-hazards-earth-syst-sci.net/8/657/2008/

24. Eftaxias, K., Athanasopoulou, L., Balasis, G., Kalimeri, M., Nikolopoulos, S., Contoyiannis, Y., Kopanas, J., Antonopoulos, G., and Nomicos, C.: Unfolding the procedure of characterizing recorded ultra low frequency, kHZ and MHz electromagnetic anomalies prior to the L'Aquila earthquake as preseismic ones – Part 1, Nat. Hazards Earth Syst. Sci., 9, 1953–1971, 2009. doi:10.5194/nhess-9-1953-2009;

25. Eftaxias, K., Balasis, G., Contoyiannis, Y., Papadimitriou, C., Kalimeri, M., Athanasopoulou, L., Nikolopoulos, S., Kopanas, J., Antonopoulos, G., Nomicos, C. Unfolding the procedure of characterizing recorded ultra low frequency, kHZ and MHz electromagnetic anomalies prior to the L'Aquila earthquake as pre-seismic ones – Part 2. Nat. Hazards Earth Syst. Sci. 10, 275–294, 2010. http://dx.doi.org/10.5194/nhess-10-275-2010.

26. Eftaxias, K., Potirakis,S.M., and Chelidze,T. On the puzzling feature of the silence of precursory electromagnetic emissions. Nat. Hazards Earth Syst. Sci.,13, 2381–2397; doi:10.5194/nhess-13-2381. 2013 .

27. Elshin Oleg, Tronin A. Andrew, Open Journal of Earthquake Research, 9, 170-180,. 2020. https://www.scirp.org/journal/ojer

28. Freund, F.T., Takeuchi, A., Lau, B.W.S. Electric currents streaming out of stressed igneous rocks – a step towards understanding pre-earthquake low frequency EM emissions. Phys. Chem. Earth 31, 389–396, 2006.

29. Gershenzon, N., Bambakidis, G. Modelling of seismoelectromagnetic phenomena. Russ. J. Earth. Sci. 3, 247–275, 2001.

30. Hadjicontis, V., Mavromatou, C., Antsyngina, T., and Chisko, K.: Mechanism of electromagnetic emission in plastically deformed ionic crystals, Phys. Rev. B, 76, 024106/1–14, 2007.

31. Hayakawa, M., Fujinawa, Y. Electromagnetic Phenomena Related to Earthquake Prediction. TERRAPUB, Tokyo. 1994.

32. Hayakawa, M. Atmospheric and Ionospheric Electromagnetic Phenomena Associated with Earthquakes. TERRAPUB, Tokyo. 1999.

33. Hayakawa, M., Molchanov, O.A. Seismo-Electromagnetics: Lithosphere- Atmosphere-Ionosphere Coupling. TERRAPUB, Tokyo, pp. 1–477. 2002.

34. Hayakawa, M., Ohta, K., Maekawa, S., Yamauchi,T., Ida,Y., Gotoh, T., Yonaiguchi, N., Sasak, H., Nakamura, T. Electromagnetic precursors to the 2004 Mid Niigata Prefecture earthquake. Physics and Chemistry of the Earth 31, 356–364. 2006. doi:10.1016/j.pce.2006.02.023

35. Hayakawa,M., Molchanov,O.A. Seismo Electromagnetics as a New Field of Radiophysics: Electromagnetic Phenomena Associated with Earthquake. The Radio Science Bulletin No 320 March 2007.

36. Ikeya, M., Takaki, S. Electromagnetic fault for earthquake lightning. Jpn. J. Appl. Phys. Part 2: Lett. 35 (3A), L355–L357. 1996.

37. Kachakhidze, M.K., Kachakhidze,N.,K., Kaladze,T.D. A model of the generation of electromagnetic emissions detected prior to earthquakes. Physics and Chemistry of the Earth, 85–86, p. 78–81, 2015. http://dx.doi.org/10.1016/j.pce.2015.02.010

38. Kachakhidze, Manana; Kachakhidze-Murphy, Nino; Khvitia, Badri; Ramishvili, Giorgi. Large Earthquake Prediction Methods. Open Journal of Earthquake Research, 8, 239-254, 2019. https://www.scirp.org/journal/ojer .

39. Khamidov, Kh.L. Assessment of strain effect of strong-motion (focus) zones of earthquakes on earth's surface displacement. Geodesy and Geodynamics 8, 34-40, 2017. www.keaipublishing.com/en/journals/geog ; http://www.jgg09.com/jweb_ddcl_en/EN/volumn/home.shtml

40. Maggipinto, Tommaso., Biagi Pier Francesco., Colella, Roberto., Schiavulli, Luigi., Ligonzo, Teresa., Ermini, Anita., Martinelli, Giovanni., Moldovan, Iren., Silva, Hugo., Contadakis, Michael., Skeberis, Christos., Zaharis, Zaharias., Scordilis, Emmanuel., Katzis, Konstantinos., Buyuksarac, Aydın., D'Amico, Sebastiano.The LF radio anomaly observed before the Mw = 6.5 earthquake in Crete on October 12, 2013. Physics and Chemistry of the Earth, Parts A/B/C.Volumes 85–86, 98-105, 2015. https://doi.org/10.1016/j.pce.2015.10.010

41. Mjachkin, V.I. Earthquake preparation processes. Moscow, "Nauka", pp.230, 1978.

42. Mognaschi, E.R. IW2GOO. On the possible origin, propagation and detectebility of electromagnetic precursors of earthquakes. Atti Ticinensi di Scienze della Terra 43, 111–118. 2002.

43. Molchanov, O. A. and Hayakawa, M.: Subionospheric VLF signal perturbations possibly related to earthquakes, J. Geophys. Res., 103, 17 489–17 504, 1998.

44. Moldovan, I.A., Constantini, A.P., Biagi, P.F., Toma Danila, D.,Moldovan, A.S., Dolea, P., Toader, V.E., Maggipinto, T.The Development of the Romanian VLF/LF Monitoring System as Part of the International Network for Frontier Research on Earthquake Precursors (INFREP). Rom. Journ. Phys., Vol. 60, Nos. 7–8, P. 1203–1217, Bucharest, 2015.

45. Onishi, Tatsuo., Parrot, Michel., Berthelier, Jean-Jacques. The DEMETER mission, recent investigations on ionospheric effects associated with man-made activities and seismic phenomena. Comptes Rendus Physique. 12 , 60–170, 2011. doi:10.1016/j.crhy.2010.11.009 .

46. Papadopoulos, G. A., Charalampakis, M., Fokaefs, A., and Minadakis, G. Strong foreshock signal preceding the L'Aquila (Italy) earthquake (Mw6.3) of 6 April 2009, 2010. Nat. Hazards Earth Syst. Sci., 10, 19–24, doi:10.5194/nhess-10-19-2010 ;

47. Politis, D., Potirakis, S. M. , Hayakawa, M. Criticality analysis of 3-year-long VLF subionospheric propagation data possibly related to significant earthquake
events in Japan. Natural Hazards, 102:47–66, 2020.
https://doi.org/10.1007/s11069-020-03910-3

48. Potirakis, Stelios M., Eftaxias, Konstantinos, Schekotov, Alexander.,Yamaguch, Hiroki, Hayakawa, Masashi.Criticality features in ultra-low frequency magnetic fields prior to the 2013 M6.3 Kobe earthquake. ANNALS OF GEOPHYSICS, 59, 3, 2016. S0317; doi:10.4401/ag-6863

49. Potirakis, S.M., Karadimitrakis, A., Eftaxias, K. 2013. Natural time analysis of critical phenomena: the case of pre-fracture electromagnetic emissions. CHAOS 23 (3117), 1–14. 2013. http://dx.doi.org/10.1063/1.4807908.

50. Potirakis, S.M., Contoyiannis, Y., Eftaxias, K., Koulouras, G., Nomicos, C. Recent field observations indicating an earth system in critical condition before the occurrence of a significant earthquake. Geosci. Remote Sens. Lett. IEEE 12, 631–635, 2015. http://dx.doi.org/10.1109/LGRS.2014.2354374 .

51. Rabinovitch, A., Frid, V., and Bahat, D.: Surface Oscillations A Possible Source of Fracture Induced Electromagnetic Radiation,Tectonophysics, 431, 15–22, 2007.

52. Reilinger, R., McClusky, S., Vernant, P., Lawrence, S., Ergintav, S., Çakmak, R., Özener, H., Kadirov, F., Guliev, I., Stepanyan, R., Nadariya, M., Hahubia, G., Mahmoud, S., Sakr, K., ArRajehi, A., Paradissis, D., Al-Aydrus, A., Prilepin, M., Guseva, T., Evren, E., Dmitrotsa, A., Filikov, S.V., Gomez, F., Al-Ghazzi, R., Karam, G., GPS constraints on continental deformation in the Africa–Arabia–Eurasia con-tinental collision zone and implications for the dynamics of plate interactions, 2006. J.Geophys. Res.BO5411. http://dx.doi.org/10.1029/2005JB004051

53. Rozhnoi, A., Solovieva, M., Molchanov, O., Schwingenschuh, K., Boudjada, M., Biagi, P. F., Maggipinto, T., Castellana, L., Ermini, A., and Hayakawa, M. Anomalies in VLF radio signals prior the Abruzzo earthquake (M=6.3) on 6 April 2009.Nat. Hazards Earth Syst. Sci., 9, 1727–1732, 2009. mwww.nat-hazards-earth-syst-sci.net/9/1727/2009.

54. Schwingenschuh, K., Prattes, G., Besser, B. P., Mocˇnik, K. ,. Stachel, M., Aydogar, O., Jernej, I. , Boudjada, M. Y., Stangl, G., Rozhnoi, A., Solovieva, M., Biagi, P. F., Hayakawa, M. and Eichelberger, H. U. The Graz seismo-electromagnetic VLF facility. Nat. Hazards Earth Syst. Sci., 11, 1121–1127, 2011. www.nat-hazards-earth-syst-sci.net/11/1121/2011/ doi:10.5194/nhess-11-1121-2011

55. Sokhadze, G., M. Floyd, T.Godoladze, R.King, E.S.Cowgill, Z.Javakhishvili, G.Hahubia, R.Reilinger. Active convergence between the Lesser and Greater Caucasus in Georgia: Constraints on the tectonic evolution of the Lesser–Greater Caucasus continental collision. Earth and Planetary Science Letters 48,1 154–161, 2018. https://doi.org/10.1016/j.epsl.2017.10.007

56. Uyeda, S., Nagao, T., and Kamogawa, M.: Short-term earthquake prediction: Current status of seismo-electromagnetics, Tectono- physics, 470, 205–213, 2009.

57. Yoshida, S., Uyeshima, M., Nakatani, M. Electric potential changes associated with slip failure of granite: preseismic and coseismic signals. J. Geophys. Res. B: Solid Earth 102 (B7), 14883–14897. 1997.

58. Yoshino, T. Low-frequency seismogenic electromagnetic emissions as precursors to earthquakes and volcanic eruptions in Japan. J. Sci. Explor. 5 (I), 121–144. 1991.

59. Zhang, X., Shen, X., Parrot, M., Zeren, Z., Ouyang, X., Liu, J., Qian, J., Zhao, S. and Miao, Y. Phenomena of electrostatic perturbations before strong earthquakes (2005–2010) observed on DEMETER. Nat. Hazards Earth Syst. Sci., 12, 75–83, 2012. www.nat-hazards-earth-syst-sci.net/12/75/2012/